\documentclass[twocolumn,showpacs]{revtex4}
\usepackage{graphicx}% Include figure files
\usepackage{dcolumn}% Align table columns on decimal point
\usepackage{bm}% bold math

\begin{document}

%\draft

\title{Spin splitting of X-related donor impurity states in an AlAs barrier. }

\author{\mbox{E. E. Vdovin$^{(1)},$} \mbox{Yu. N. Khanin$^{(1)},$}\mbox{L. Eaves$^{(2)},$} \mbox{ M. Henini$^{(2)},$} \mbox{and  G. Hill$^{(3)}$}}

\address
{$^{1}$Institute of Microelectronics Technology and High Purity
Material, Russian Academy of Sciences,Chernogolovka, Moscow
District, 142432 Russia}
\address
{$^{2}$School of Physics and Astronomy, University of Nottingham,
Nottingham, NG7 2RD, UK. }
\address
{$^{3}$Department of Electronic and Electrical Engineering,
University of Sheffield, Sheffield, S3 3JD, UK.}

\begin{abstract}
We use magnetotunneling spectroscopy to observe the  spin splitting of the ground state
of an X-valley-related Si-donor impurity in an AlAs barrier. We determine the absolute
magnitude of the effective Zeeman spin splitting factors of the impurity ground state to be
g$_{I }$= 2.2 $\pm $ 0.1.  We also investigate the spatial form of the electron wave function of the donor
ground state, which  is anisotropic in the growth plane.
\end{abstract}

\maketitle

\subsection*{1. Introduction}

The effect of spin on electronic transport has recently attracted much interest both from an
applied and from a fundamental point of view \cite{Loss,Hanson}. Resonant electron tunneling through the
discrete states of self-assembled semiconductor quantum dots (QDs) and the physically similar
bound states of impurities is a promising means of studying spin-resolved transport. Such
tunneling experiments have been used to observe directly the spin splitting of the bound
electronic states of shallow impurities within a GaAs quantum well \cite{Sakai,Deshpande,Koneman} or of electrons in InAs
QDs \cite{Thornton,Hapke,Hapke2}, and to measure directly the absolute value of the $g$-factor of these zero-dimensional
states. In addition, they have been used to probe at a mesoscopic level the spin dependence of the
local density of states \cite{Konig}. Resonant tunneling studies have also been performed on single barrier
GaAs/AlAs/GaAs heterostructures. AlAs is an indirect gap material with the minima of the
conduction band at the X points of the Brillouin zone, whereas in GaAs the minimum is at the $\Gamma $
point. There have been several earlier studies of tunneling through X-valley states  \cite{Mendez,Carbonneau,Finley,Teissier,Smith,Im,Im2},
including investigations of the tunneling through donor states associated with the X conduction
band minima \cite{Fukuyama,Khanin,Itskevich,Vitusevich,Khanin2,Gryglas,Khanin3}.

In this paper, we report the observation of Zeeman spin splitting of the ground state of an
Si-donor impurity embedded in an AlAs tunnel barrier for the orientation in which the magnetic
field is applied in the plane of the barrier, i.e. perpendicular to the direction of the electron tunnel
current.  This state is associated with the anisotropic X-conduction band minima of AlAs. The Si
atoms, which are located substitutionally on Al sites, diffuse during growth into the AlAs barrier
from adjacent GaAs layers which are $\delta $-doped with Si. We  measure the effective $g$-factor of
the zero-dimensional state and obtain the absolute values of the effective spin splitting factor
components $g$ of between 2.1 and 2.2. In addition, magnetotunneling spectroscopy provides us
with information about the spatial form of the wave function of an electron bound in the X-
related donor state. Our measurements indicate that the wave function has a biaxial symmetry in
the growth plane, with axes corresponding to the main crystallographic directions of the (001)-
epilayers.

Let us first briefly review the previously-studied problem of tunneling through isolated
donor impurities in the GaAs quantum well (QW) of large area GaAs/(AlGa)As double barrier
resonant tunneling diodes. In this case  the donors form localized ($\sim $10 nm) hydrogenic bound states
associated with the $\Gamma $-conduction band minimum of the GaAs QW \cite{Sakai2}. These states are located
at an energy of $\sim $10 meV below the bottom of the lowest energy subband of the QW. Under an
applied bias the tunnel current exhibits a rapid increase when the impurity state aligns with the
Fermi level in the negatively-biased electron emitter layer. In general, there are many
impurities giving rise to multiple, overlapping steps in the current-voltage characteristics, and
these multiple peaks can be resolved in the current-voltage characteristics of small area mesa
samples \cite{Gryglas}. Similarly, in our previous work, the resonant tunneling of electrons through
individual X-related donor impurity states of a single, relatively thin, 5 nm AlAs barrier (with an
X conduction band quantum well) appeared as partially resolved fine structure in a broad
resonance, associated with the ensemble of donors \cite{Khanin2}. This fine structure arises because the
donors are located in different atomic planes of the AlAs and the spectrum of donor states is determined predominantly
by the dependence of the binding energy on the position of the donor in the barrier.The influence of the random
variations of the electrostatic potential on the energies of the donor impurity in this case is insignificant.

In contrast, for the experiments described here, the donors are randomly located in a
relatively thick, 11.2 nm, barrier, so the influence of the random electrostatic potential is
considerable. The essential role of the random variations of the electrostatic potential in this case
is associated with the presence of the $\delta $-doped layers near the barrier \cite{McDonnell} and the slow
dependence of the binding energy of donors on their position in the thick barrier  \cite{Weber}. As a
result, resonant tunneling of electrons through the donor states gives rise to a series of sharp,
well-resolved peaks in the $I(V) $  curves. We ascribe each peak to tunneling through a single or
very small number of individual donor states. This allows us to observe spin splitting of the
donor resonances and to determine the g-factor of the zero-dimensional states directly.

\subsection*{2. Samples}

A schematic diagram of our device is shown in Fig. 1. The active part of our
samples comprises a single 11.2 nm thick AlAs barrier which is sandwiched
between two accumulation layers formed by two $\delta $-doped layers with
Si-concentration of 3 $\times $ 10$^{11}$ cm$^{ - 2}$, located at a
distance of 2.8 nm from each side of the barrier. A two-dimensional electron
gas (2DEG) forms in each accumulation layer at zero bias. The AlAs layer was
not intentionally doped, but donor impurities are present in the AlAs due to
diffusion of Si into the barrier from the $\delta $-doped layers. The
calculated $\Gamma $ and X band profiles of the active part of our device at
zero bias are shown in Fig.1. The heterostructures was grown by molecular
beam epitaxy on a (001)-oriented, Si doped n$^{ + }$-type GaAs wafer (
N$_{d}$ = 2$\times $10 $^{18}$ cm $^{ - 3}$ ) at a temperature of 550 $^{0}$
C. The detailed layer composition of the heterostructure, in order of growth
on the substrate, is as follows: a Si-doped, 0.5 $\mu $m-thick GaAs buffer
layer (N$_{d}$ = 2$\times $10$^{18}$cm$^{ - 3})$, a 60 nm-thick GaAs layer
(N$_{d}$ = 3$\times $10$^{17}$cm$^{ - 3})$, a 21.6 nm-thick undoped GaAs
layer; a 5.6 nm-thick undoped Ga$_{0.9}$Al$_{0.1}$As layer, a 28 nm-thick
undoped GaAs layer, a Si $\delta $-doped layer with concentration of 3 x
10$^{11}$ cm$^{ - 2}$, a 2.8 nm-thick undoped GaAs layer, a 11.2 nm-thick
AlAs barrier layer; a 2.8 nm-thick undoped GaAs layer, a Si $\delta $-doped
layer with concentration of 3 x 10$^{11}$ cm$^{ - 2}$, a 28 nm-thick undoped
GaAs layer, 5.6 nm-thick undoped Ga$_{0.9}$Al$_{0.1}$As layer, a 21.6 nm-thick undoped
GaAs layer, a 60 nm-thick GaAs layer (N$_{d}$ = 3$\times $10$^{17}$ cm$^{ -
3})$, and a 0.5 $\mu $m-thick, GaAs cap-layer (N$_{d}$=2$\times
$10$^{18}$cm$^{ - 3})$. Ohmic contacts were made by deposition and annealing
of AuGe/Ni/Au layers. Mesa structures, with diameter between 50 $\mu $m and
200 $\mu $m, were fabricated by chemical etching.

\subsection*{3. Experimental}

Tunnel current measurements at constant applied voltage with magnetic field B applied
parallel to the current (i.e. perpendicular to the 2DEG) reveal Shubnikov-de-Haas (SdH)-like
oscillations  \cite{Chan}. Close to zero applied bias, analysis of the SdH-like oscillations gives a value of
n$_{s}$ = 3.27 $\times $ 10$^{11}$ cm$^{ - 2}$ for the sheet density of the two 2DEG layers. Figure 2(b) shows the low
temperature (4.2K) current-voltage, $I(V) $ , characteristics at low bias voltages for a typical device
$i$ which exhibit sharp peaks in the current over voltage range from 10 to 60 mV. This peak
structure is observed to be sample-specific, but for a given sample it is exactly reproducible from
one voltage sweep to another. The peaks are reproducible even after thermal cycling of the
sample, except for a small voltage shift ($\sim $ a few mV). We ascribe the peaks in current to single
electron tunneling through individual, zero dimensional Si-donor states in the AlAs barrier.
Similar features have been observed and reported previously in large area double barrier RTDs
 \cite{Sakai,Koneman,Sakai3} and attributed to tunneling through individual tunneling channels due to zero-
dimensional states. Increasing the voltage across the device moves the energy of the donor state
relatively to the Fermi level of the 2DEG which acts as an emitter for the tunnelling electrons
(Fig. 2(a)). Tunneling occurs as the donor state crosses the Fermi level of the 2DEG, and stops
when the donor state is brought below the 2DEG subband edge.

Figure 2 show $I(V) $ at 0T and 8 T with magnetic field oriented perpendicular to the current
direction. In a magnetic field the ground state of an Si-donor impurity splits into two spin energy
levels given by

\begin{equation}
E _{Si}=g_{I}  \mu _{B}Bm_{s} (m_{s}=\pm 1/2),   \label{1}
\end{equation}

where  $g_{I}$ is the $g$-factor of the Si-donor impurity.  This opens up two separate channels for
electrons from the 2DEG to tunnel into, and we therefore see separate peaks in $I(V)$ due to
electrons tunneling through each of these spin energy levels. In a magnetic field applied
perpendicular to the current (i.e. parallel to the 2DEG) the 2DEG emitter becomes partially spin
polarized, due to energy splitting of the Fermi energies of the two spin species, as shown
schematically in Figure 2(a). Due to the slow tunneling rate from the 2DEG, the two spin species
in the 2DEG should be in thermal equilibrium, and so the chemical potential of each is the same.
Therefore, there is an energy difference between the subband edge of the spin species, equal to
the spin-splitting $g_{2D}$\textit{$\mu $}$_{B}B$, where $g_{2D}$ is the $g$-factor of electrons
in the 2DEG emitter.  Resonant tunneling occurs when an impurity spin level crosses the Fermi level of the 2DEG. We assume
that spin is conserved during the tunnelling process.  For each spin we see a separate onset of
tunneling and the voltage difference between the position of the onsets \textit{($\Delta $V}$_{peak})$ is proportional to
the energy difference, \textit{$\Delta $E}$_{ Si }=g_{I}$\textit{$\mu $}$_{B}B$, obtained from Eq.(1).

Figure 3(a) shows in detail the behavior of the first current peak of device \textit{ii} at 4.2 K in
different magnetic fields applied perpendicular to the current direction from B=0 T to 8 T, in
steps of 0.5 T. Figure 3(b) shows the voltage separation between the corresponding two spin-
split peaks which increases linearly with magnetic field strength, as expected for a Zeeman
effect. Because of the finite widths of the current peaks, it is not possible to resolve the splitting
for magnetic fields less than 5 T. The best fit line to the data closely intersects $\Delta $V=0  at B=0 T
and has a slope $g_{I }$\textit{$\mu $}$_{ B}$ / $f$, where {$\mu $}$_{ B}$  is the Bohr magneton, $g_{I}$ is
the effective gyromagnetic ratio of the impurity with the magnetic field perpendicular to the current
(i.e. perpendicular to the growth direction of the quantum well), and $f$ is the so-called electrostatic leverage factor. The
temperature dependence of the current onset allows us to determine the electrostatic leverage
factor $f$  \cite{Deshpande,Thornton}. Figure 4 shows the temperature dependence of the onset of a typical resonant
current features. We deduce a value of 0.44 for the electrostatic leverage factor by fitting the
Fermi-Dirac function to the form of the measured low bias onset of the peak in current at various
temperatures, using the procedure described in reference \cite{Thornton}. A similar value of $f$ is obtained
from self-consistent Poisson-Schr\"{o}dinger calculations: these indicate that, over the bias range of
interest (0-100 mV), the leverage factor $f$ for an electron tunneling from the emitter into an
impurity located at the center of AlAs barrier varies slightly from 0.44 to 0.42 eV per volt of
applied bias. The uncertainty in our value of $g$ is determined by the error in the leverage factor.
Note also from Figure 4 that the resonant peak in $I(V)$  is strongly enhanced as the temperature is
reduced.  The enhancement, which may be related to a many-body Fermi energy singularity
effect \cite{Geim}, tends to improve the resolution of the spin-split peaks.

We have measured the  splitting of several peaks in $I(V) $ and find that the different
impurity-related peaks give values of $g_{I}$ in the range from 2.1 to 2.22. Our value of the $g$-factor
of the X-related impurity states in AlAs is of a larger absolute value than reported in another
tunneling experiment \cite{Vitusevich}, where $g$=0.34.  However, for the experiment described here, the
donors are located in a relatively thick, 11.2 nm, AlAs layer, whereas the localized state
investigated in Ref. \cite{Vitusevich} was embedded in a narrow 2 nm AlAs barrier. This value of $g$=0.34 is
quite different from that for the X-valley electrons in bulk AlAs. The $g$-factor for electrons in
bulk AlAs expected from theoretical calculations is 1.9 \cite{Roth}, and the $g$-factor of electrons in bulk
Al$_{0.8}$Ga$_{0.2}$As has been measured by electron paramagnetic resonance to be 1.96 \cite{Bottcher}. Also, van
Kesteren \textit{et al.} have reported a value of $ \approx $ 1.97 for electrons in AlAs QWs based on optically
detected magnetic resonance experiments on AlAs-GaAs superlattices \cite{Kesteren}. This difference in
values may be due to the complex nature of the X-related donor impurity states in AlAs barrier
\cite{Vitusevich}. In low-dimensional heterostructures, it is known that the value of the $g$ factor can be
modified from its bulk value, due to quantum confinement effects and because the electron wave
function contains contributions from the different materials which make up the structure
\cite{Deshpande,Snelling,Ivchenko}. Calculations show that $g$ is a strong function of the quantum well width
\cite{Kiselev}, and the rather wide X minima quantum well of our AlAs barrier gives a $g$-factor of the X-donor in our
experiment which is close to the $g$-value of bulk AlAs $g \approx $2, since the modification of the band
structure due to the quantum confinement is fairly small.

We also studied the $g$-factor dependence for magnetic field applied along different
crystallographic axes in the (001) growth plane. In contrast to the case of single-electron
tunneling through the localized states of InAs quantum dots \cite{Hapke2}, to within the resolution of our
experiment, the spin splitting of the X-related donor impurity in AlAs was isotropic with respect
to the angle of the in-plane magnetic field.

We now consider the magnetic field dependence of the amplitude of the tunnel current
through the X-related donor impurities as a function of magnetic field B applied perpendicular to
the direction of tunneling. We attribute the general fall in amplitude of both spin-split
components with increasing B (see Fig. 5(a)) to a well-established effect that can be understood
in term of a single-particle model for electron tunneling in the presence of magnetic field \cite{Sakai2,Vdovin,Patane}.
Let \textit{$\alpha $}, \textit{$\beta $} and $Z $ indicate, respectively, the direction of $B$, the direction normal to $B$ in the
growth plane ($X$, $Y)$, and the normal to the tunnel barrier, respectively. When an electron tunnels
from the emitter accumulation layer into the impurity state in the barrier, it acquires an additional
in-plane momentum given by

\begin{equation}
k_\beta = eB\Delta s / \hbar   , \label{2}
\end{equation}

where $\Delta s$  is the effective distance tunnelled along $Z $( $\sim $ 8  nm for our device ). This gives the
increased momentum along \textit{$\beta $}, which is acquired by the tunnelling electron due to the action of
the Lorentz force.

The applied voltage allows us to tune resonantly to the energy of a particular impurity state.
Thus, by measuring the variation of the tunnel current with $B$, we can determine the size of the
matrix element that governs the quantum transition of an electron as it tunnels from a state in the
emitter layer into an impurity state.

In order to analyse the results of our experiment, we express the tunneling matrix element $M$
in terms of the Fourier transforms $\Phi _{i(f)} (k)$   of the conventional real space wave functions,
according to the relation  $M = \int_k {\Phi _i (k - k_\beta } )\Phi _f (k)dk$, and express the tunneling current as
$I \quad \sim  \left| M \right|^{2}$ \cite{Sakai2,Patane}. Here the subscripts $i$ and $f$ indicate the initial
(emitter) and final (Si-impurity) states of the tunnel transition. Relative to the strong spatial confinement in the zero dimensional
impurity state, the initial state in the emitter is essentially unconfined - i.e. it behaves like a free
particle. Hence, in $k$-space  $\Phi _i (k)$  corresponds to a sharply peaked function with a finite value
only close to $k$= 0. Since the tunnel current is given by the square of the matrix element
involving  $\Phi _i (k)$ and the Fourier transform of the Si-donor wave function, $\Phi _{Si} (k)$ , the narrow
spread of $k$ for $\Phi _i (k)$  allows us to investigate the form of  $\Phi _{Si} (k)$ by varying $B$ and hence $k$,
according to equation 2. Thus by plotting $I(B)$ for a  \textit{particular} direction of $B$ we can measure the
dependence of $\left| {\Phi _{Si} (k)} \right|^2$  along the $k$ -direction perpendicular to $B$. Then, by rotating $B$ in the
plane ($X, Y)$ and making a series of measurements of $I(B)$ with $B$ set at regular intervals
($\Delta \theta _{ }\sim $ 15\r{ })  of the rotation angle $\theta$ , we obtain a full spatial profile of
$ \left| {\Phi _{Si} (k_X ,k_Y )} \right|^2$. This represents the projection in $k$-space of the probability
 density of a given impurity electronic state peak.

Typical experimental data for sample \textit{ii}, of the variation of the current at peak A with the
direction of magnetic field at 8 T, are shown as polar plots in Fig.5 (b) and (c). A maximum
current modulation $\Delta I/I $ of about $\sim $ 23{\%}  is observed, with clear, twofold anisotropy observed for
the two split peaks A$^{ - }$ and A$^{ + }$. The anisotropy of each observed peaks has a similar magnitude
and orientation. In Fig.5, 0$^{o}$ corresponds to the [110] direction, so the principal axes for the anisotropy are
oriented along [100] and [010] directions. This result shows that the wave function shape of the X-related donor
impurity in AlAs barrier is anisotropic in the growth plane, with the wave function probability density elongated
along the direction [100] in real space. This is in contrast to the case of Si donor states in a GaAs
quantum well, where the electron wave function has circular symmetry in the growth plane, as
expected for 1$s $ donor ground state \cite{Patane}. We suggest that this anisotropy may be related to
the anisotropy of the effective mass of the electrons in the X conduction band minima of AlAs,
though this point requires further theoretical analysis.

In conclusion, we have used magnetotunneling spectroscopy to observe the  spin
splitting of the ground state of Si donor impurities in an AlAs tunnel barrier. These states are
associated with the X-conduction band minima of AlAs. We determine the absolute magnitude
of the anisotropic effective magnetic spin splitting factors, $g$, for these states to be 2.1 $\pm
$ 0.1. In addition, we use magnetotunneling spectroscopy to investigate the spatial form of the wave
function of the X-related donor impurity. The wave function of electrons bound to an X-related
donor has a biaxial symmetry in the growth plane, with axes corresponding to the main
crystallographic directions.

The work is partly supported by RFBR (03-02-17693) and EPSRC (UK). EEV gratefully
acknowledge support from the Royal Society. The authors thank Y.V. Dubrovskii and K.A.
Benedict for useful discussions, and V.V. Belov for technical assistance.

\begin{figure}\includegraphics{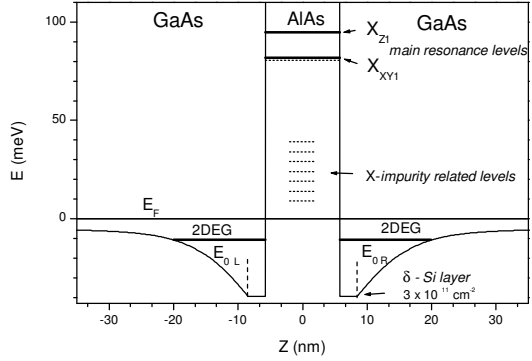}
\caption{Calculated conduction band profile of the active part of
the tunnel structure at zero applied voltage. The figure shows the
positions of the Fermi level $E_{F}$, the quantised GaAs
accumulation layer subbands $E_{0L}$ and $E_{0R}$, and the
size-quantized levels of the $X_{Z}$ and $X_{XY}$ subbands in the
AlAs barrier. The energy positions of the X-impurity related levels
in AlAs are also shown.}
\end{figure}

\begin{figure}\includegraphics{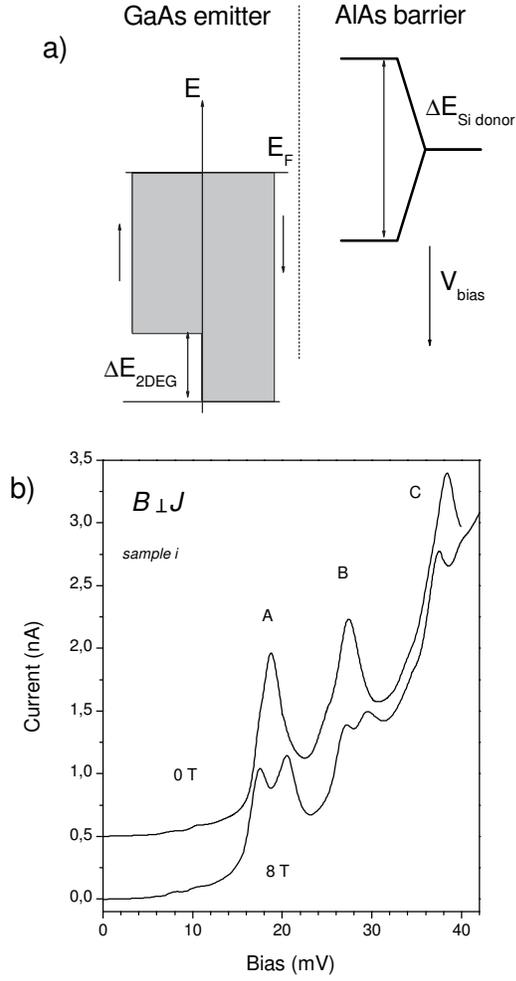}
\caption{(a) Schematic diagram of the spin splitting of the Si donor
state in the AlAs and partial spin polarization of the 2DEG in a
magnetic field applied perpendicular to the current. The effect of
applying a voltage across the device is move the energy donor energy
levels down relative to the Fermi level of the 2DEG. (b) $I(V)$
characteristics at 4.2 K of sample \textit{i} at 0 T and 8 T for
magnetic field applied perpendicular to the current. The curves are
offset for clarity. The characteristics show sharp peaks in the
current due to tunneling through discrete X-impurity states, each of
which split in a applied magnetic field.}
\end{figure}

\begin{figure}\includegraphics{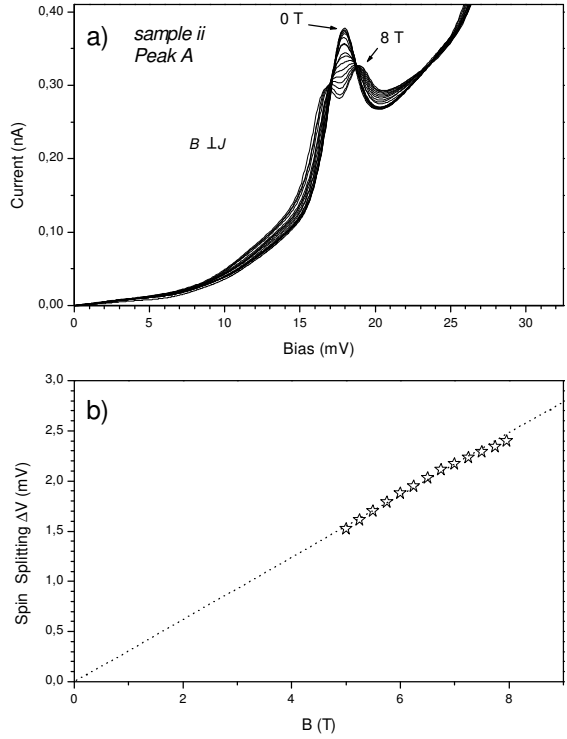}
\caption{(a) Evolution of the spin-split peaks in $I(V)$ the sample
\textit{ii} at 4.2 K in various magnetic fields perpendicular to the
current from B=0 T to 8 T, in steps of 0.5 T. (b) The measured spin
splitting versus magnetic field. The dashed lines are linear fits to
the data.}
\end{figure}

\begin{figure}\includegraphics{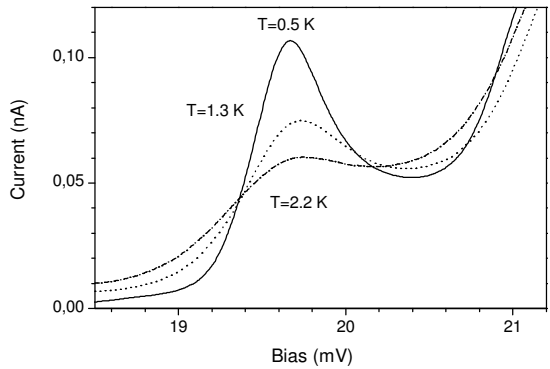}
\caption{$I(V)$ characteristics of the first current peak of the
typical samples at different temperatures showing the Fermi-level
broadening. }
\end{figure}

\begin{figure}\includegraphics{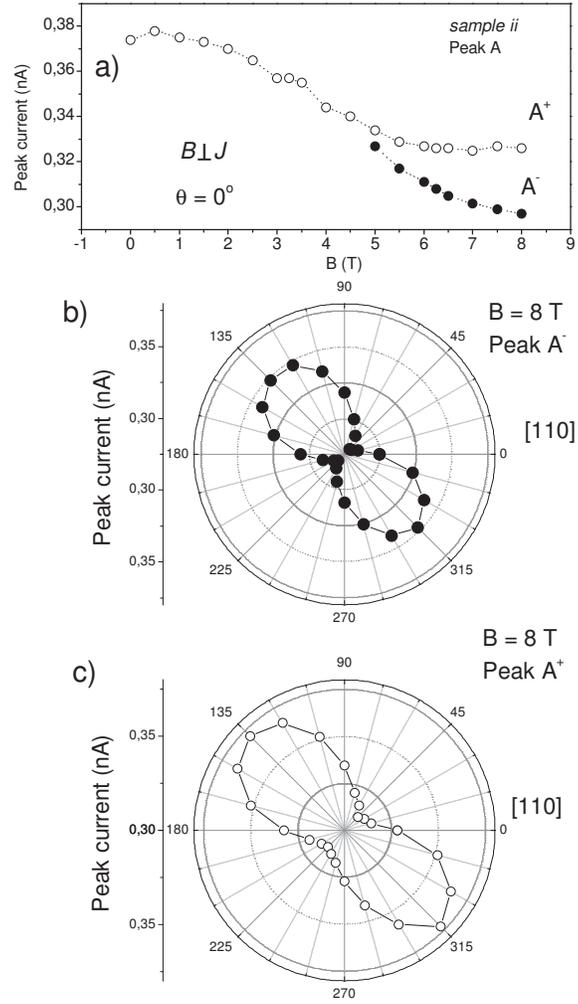}
\caption{(a) Amplitude dependence of the peak A of sample
\textit{ii} vs  magnetic field  \textbf{\textit{B}} applied parallel
to a [110] direction in the plane of the quantum well. Polar plot of
the change in peak current vs in-plane magnetic field direction for
the peaks A+ (b) and A- (c) at 8 T. }
\end{figure}

\end{document}